\documentclass[runningheads]{llncs}
\usepackage{graphicx}
\usepackage{todonotes}
\begin{document}
\title{XR Hackathon Going Online: Lessons Learned from a Case Study with Goethe-Institut}
\titlerunning{XR Hackathon Going Online: Lessons Learned}
%
\author{Wiesław Kopeć\inst{1,2}\orcidID{0000-0001-9132-4171} \and
Kinga Skorupska\inst{1,2}\orcidID{0000-0002-9005-0348} \and
Anna Jaskulska\inst{1,3}\orcidID{0000-0002-2539-3934} \and
Michał Łukasik\inst{1}\orcidID{0000-0001-5463-2933} \and
Barbara Karpowicz\inst{1}\orcidID{0000-0002-7478-7374} \and
Julia Paluch\inst{1}\orcidID{0000-0002-7657-7856}\and
Kinga Kwiatkowska\inst{1}\orcidID{0000-0002-2957-3975} \and
Daniel Jabłoński\inst{1}\orcidID{0000-0002-3551-0970} \and
Rafał Masłyk\inst{1}\orcidID{0000-0003-1180-2159}}
\authorrunning{Kopeć et al.}
\institute{Polish-Japanese Academy of Information Technology \and
SWPS University of Social Sciences and Humanities \and Kobo Association
}
\maketitle              
\begin{abstract}
In this article we report a case study of a Language and Culture-oriented transdisciplinary XR hackathon organized with Goethe-Institut. The hackathon was hosted as an online event in November 2020 by our University Lab in collaboration with Goethe-Institut as a follow-up to our previous co-organized event within our research group Living Lab. We have improved the formula of the event based on lessons learned from its previous edition. First, in one of the two hackathon tracks we provided the participants with a custom VR framework, to serve as a starting point for their designs to skip the repetitive early development stage. In cooperation with our partner, Goethe-Institut, we have also outlined best modern research-backed language-learning practices and methods and gathered them into actionable evaluation criteria.

\end{abstract}

\begin{figure}
\includegraphics[width=\textwidth]{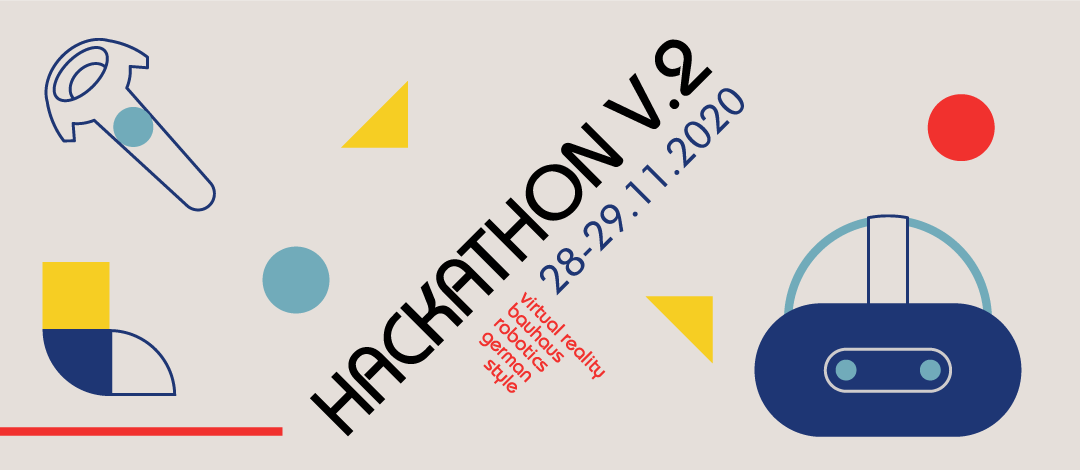}
\caption{Our social media cover image for the VR Hackathon} \label{vrcover}
\end{figure}

\section{Introduction and Related Works}

Hackathons have great potential for sparking inspiring ideas and collaborations and they provide great informal learning opportunities for students \cite{nandihackathons,mit2019hackathon,fowler2016informal} and other participants, for example older adults, who may join hackathons as experts \cite{kopec2018hackolder} - however, this massive potential is rarely realized in full. According to a survey of 150 hackathon participants \cite{briscoe2014digital}, they do not participate because they want to build a product (26\%), rather they do it for learning (86\%) and networking (82\%), especially if it is a corporate event, where the primary motivation, both of the organizers and attendees is finding employment \cite{Ru2020HackathonWorkforceDev}. Therefore, hackathons ought to be constructed in a more user-centric, or participant-centric manner, with these predominant motivations in mind, especially in learning environments. For this reason, in this event organized a year we tried an approach different from our previous practice.

The online event we organized was a follow-up to an event co-organized with Goethe-Institut a year prior, which we described in a case study at CHI 2021 \cite{kopechackchi2021} together with a rich set of insights based on the way it was organized, conducted as well as the final projects and participants' impressions. As project lifespan, and thus learning, after hackathons is limited because of motivation, follow-up, project understanding, team-composition, technologies used as well as a myriad of other factors \cite{whathappenstoprojects2020} we decided to address some of these issues in this second edition. 

First of all, we have built a clear set of guidelines and instructions in collaboration with our domain-specific expert partner institution - the Goethe-Institut. The same guidelines served to evaluate the projects after their completion. Next, keeping the learning and inclusivity in mind, we have enabled the participants to request features and changes to the hackathon formula prior to its launch. This resulted in lowering a barrier to entry, as the potential participants requested an additional track to enable them to participate without the requirement of having a functional VR headset at home.

Therefore, we added the second track devoted to AR projects, without any explicit hardware or framework but with a suggested broad pro-environmental theme. In addition to this track, we had the default VR track we have previously planned, which used our pre-fabricated custom VR framework to enable the participants to easily skip the basic and repetitive programming requirements and make the choice of technologies and the related trade-offs a non-issue \cite{hackingeventsprojectdev2017}. Finally, we provided opportunities to active and willing participants to develop the projects further under the mentorship of experts in the field of XR development. Such extended, post-event mentorship, going beyond even the active participation of the mentor in the teams' decision making processes \cite{Nolte2020HackathonMentoring}, can be a valuable asset to the participants. The opportunity to pursue follow-up work is well-aligned with participants' prevailing motivation connected to learning new things \cite{briscoe2014digital}.

Therefore, in this case study, which falls well within the recent trends of research-oriented hackathons \cite{10yofresearch2020}, we report the analysis, further insights and lessons learned from organizing a language and culture-oriented XR hackathon with Goethe-Institut. We refer to the benefits of using a provided technological framework as well as an extensive set of expert guidelines and evaluation criteria, both of which solved some of the problems we have faced in the previous edition of the hackathon.

\section{Online XR Hackathon Case Study Method and Tools}

\begin{figure}
\includegraphics[width=\linewidth]{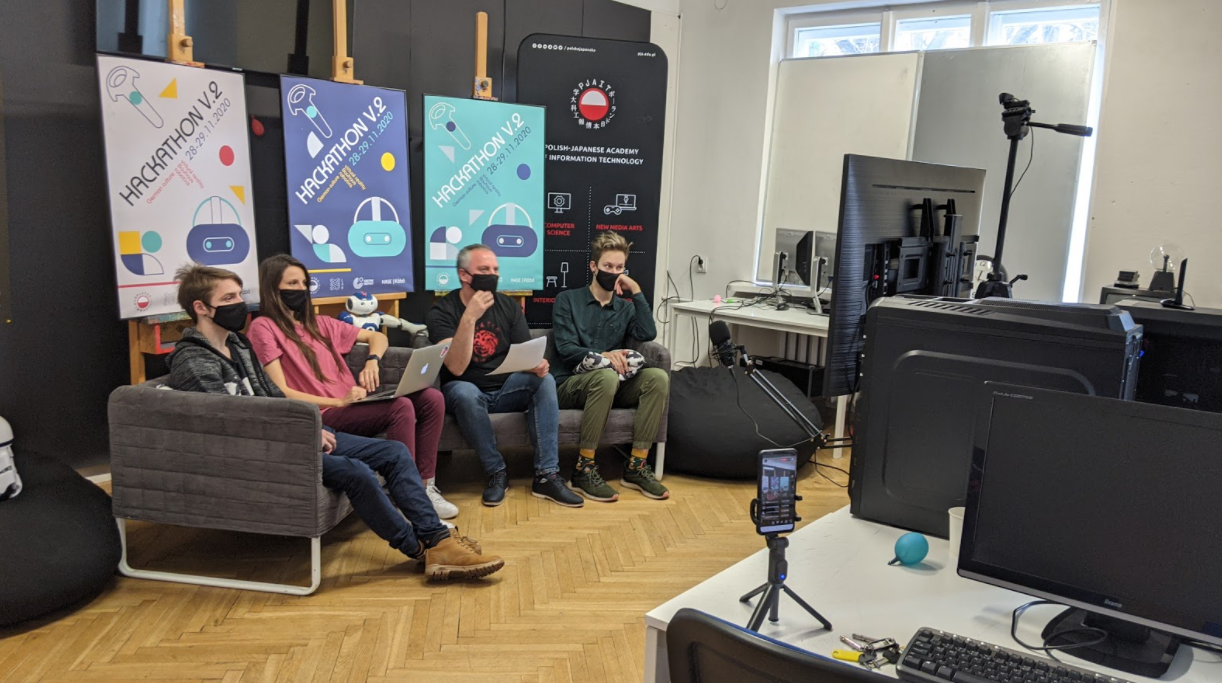}
\caption{Part of the organizing team during the hackathon's live streamed kick-off meeting} \label{teamsteam}
\end{figure}

\subsection{Key information}

The hackathon was hosted as an online event on the 28-29.11.2020 by the XR Lab of the Polish-Japanese Academy of Information Technology in collaboration with Goethe-Institut and Kobo Association as a follow-up to our previous co-organized event \cite{kopechackchi2021}. To launch and end the event as well as communicate key information we used MS Teams, which was familiar to our students their online classes were conducted on the same platform. During the hackathon participants communicated with each other and with mentors using a dedicated Discord server. Additionally, the launch and end of the hackathon, including final project presentations, were broadcast using Instagram stories.

\begin{figure}[h!]
\centering
\includegraphics[width=\linewidth]{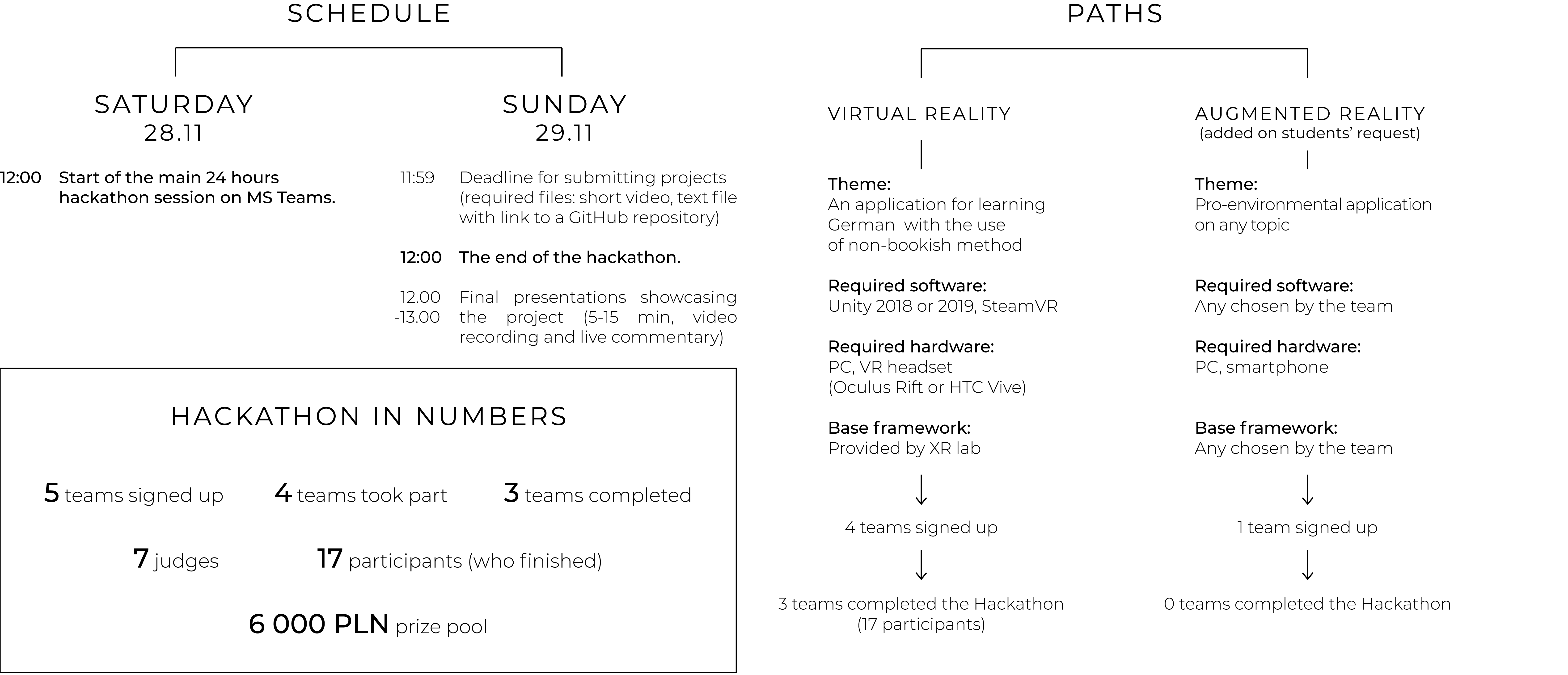}
\caption{Schematic overview of the online hackathon} \label{schematic}
\end{figure}

\subsection{Organization}

\subsubsection{AR track vs VR track with a Starting Framework}
The hackathon allowed for a choice of two technology tracks - one for AR,
and one for VR. There were clear differences between them, as the AR track had a lower barrier to entry (no need to have a VR headset) and allowed the participants to use any technology of their choice. The project topic in this track had to fall within a broader pro-environmental theme. In the VR path the participants' task was to use a custom VR Framework, programmed by our XR Lab, to create an application for learning German with the use of modern immersive methods. The provided framework was dedicated for the Unity environment and consisted of the following parts:
\begin{itemize}
\item Scenes: MainMenu (interactive about and credits board, a player tutorial for teleportation, interacting with objects (doors, grabbing) and instructions for creating new scenes.
\item Prefabs: Avatar (character animations for idle and talking) and Player (containing everything needed to navigate in VR, including hand scripting)
\item Scripts: ControllerHints, HandManager (controllers), PickupManager (interacting with grabbable objects), Interactable (class for objects one can interact with using the laser), HighlightChanger (to highlight objects which are pointed at), HighlightMode, HighlightColors, Grabbable (for objects that can be picked up with the hand or the laser) and TeleportArea (to map where the player can teleport to)	
\end{itemize}

The framework was accompanied by a comprehensive, 12-page manual outlining key functionalities, requirements and instructions for installing the development environment components, such as SteamVR, Unity 2018 or 2019 and the framework repository files.

\begin{figure*}
\centering
\includegraphics[width=350px]{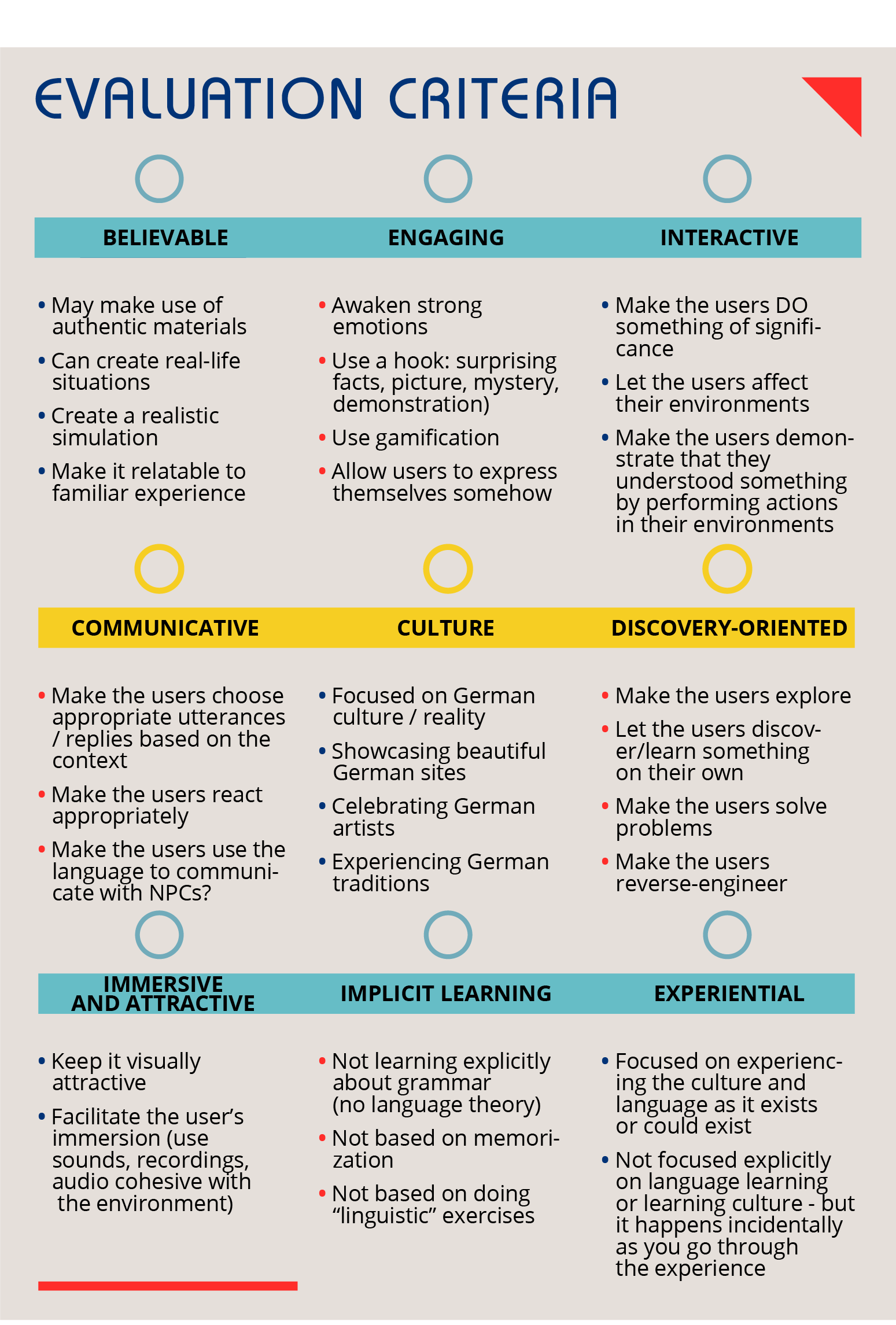}
\caption{Criteria used to evaluate the projects which were co-created by the language, art and IT experts from the jury and validated by Goethe-Institut prior to the hackathon.} \label{critnew}
\end{figure*}

\begin{figure*}
\includegraphics[width=\textwidth]{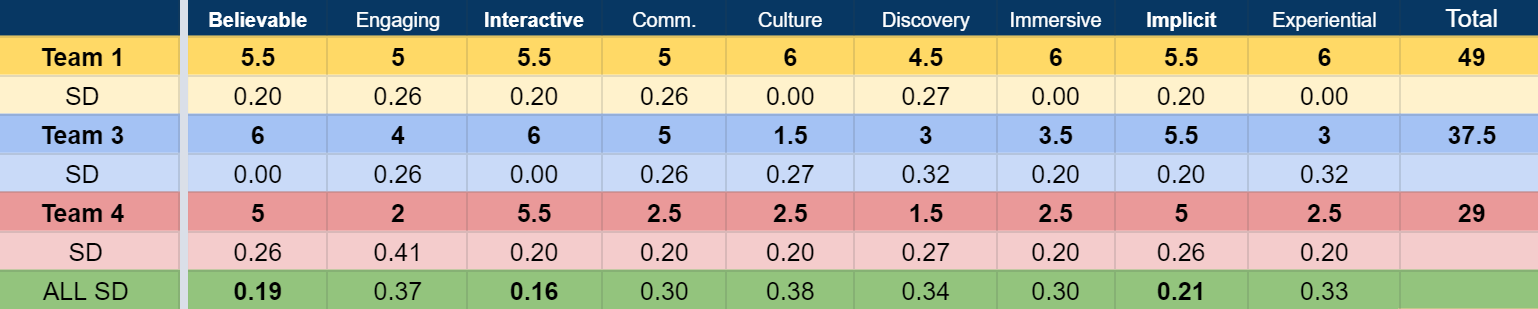}
\caption{Results of the jury evaluation of the team projects broken into specific criteria.} \label{scores}
\end{figure*}

\subsubsection{Project Presentation}

The projects were submitted to each team's own folder on Google Drive. Each submission was to include a short video recording presenting the effect of work (gameplay, assets, features etc) as well as a text file with a link to the project's repository on GitHub. The on-air 5 to 15-minute presentation of the project, the video as well as live commentary by the participating teams.

\subsubsection{Judging Criteria and Jury}

For this hackathon, as the focus of our default track was culture- and language-oriented we have crated an extensive set of evaluation criteria with a description of each one, as shown in Fig \ref{critnew}. Seven judges, experts in different areas including languages, research, IT, VR-development and art, could grant 0, 0.5 or 1 points to each project in each criterion depicted in \ref{critnew}. Additionally, each judge had the ability to grant up to 2 additional points to the projects they liked the most, provided they created their own criteria to justify their choice. The judging process was constructed this way to test the evaluation criteria chosen for this activity, as well as to enable the judges to form their own criteria, which could replace some of the existing criteria with low predictive value, or be added to the evaluation process in the future events.

\section{Results}

\subsection{Teams and Tracks}

The event was attended by 4 teams - including university and high school students. Each team consisted of at least 1 programmer and 1 artist to provide team diversity and transdiciplinarity, as per best practices \cite{whathappenstoprojects2020}. The minimum number of people in a team was 3 and the maximum 6. Five teams signed up for the Hackathon, four took part in the competition, three of which completed the project with participant count of 6, 6 and 5 in each of them. Team number 2 resigned during the competition, as they could not finish the project to a satisfactory degree. All of the teams that successfully completed the hackathon were from the VR track. Here, one limitation of the online event was connected to granting appropriate space and time for ice-breaking activities and team formation facilitated by social preferences of the participants \cite{communitysocialteamformation2016}.

\subsection{Judging Criteria}
The results of the evaluation can be seen in Fig \ref{scores}. The categories of Believable, Interactive and Implicit, as explained in Fig \ref{critnew}, failed to produce sufficiently different scoring results, with the standard deviation between all projects and all judges for these criteria at 0.19, 0.16 as well as 0.21. Similar results were obtained from interviews with the judges after the judging process, as they remarked that these criteria were either redundant (interactive, as VR by default is interactive), too general (believable) or easily achievable, thanks to prior instructions (Implicit Learning). Other criteria that the judges have specified themselves were: graphics (4 times), innovative idea (3 times) and one time for each: mood, humour, functionalities, effects and potential.

\subsection{Winning teams and outcomes}

\subsubsection{First place}

\begin{figure*}[h!]
\includegraphics[width=\textwidth]{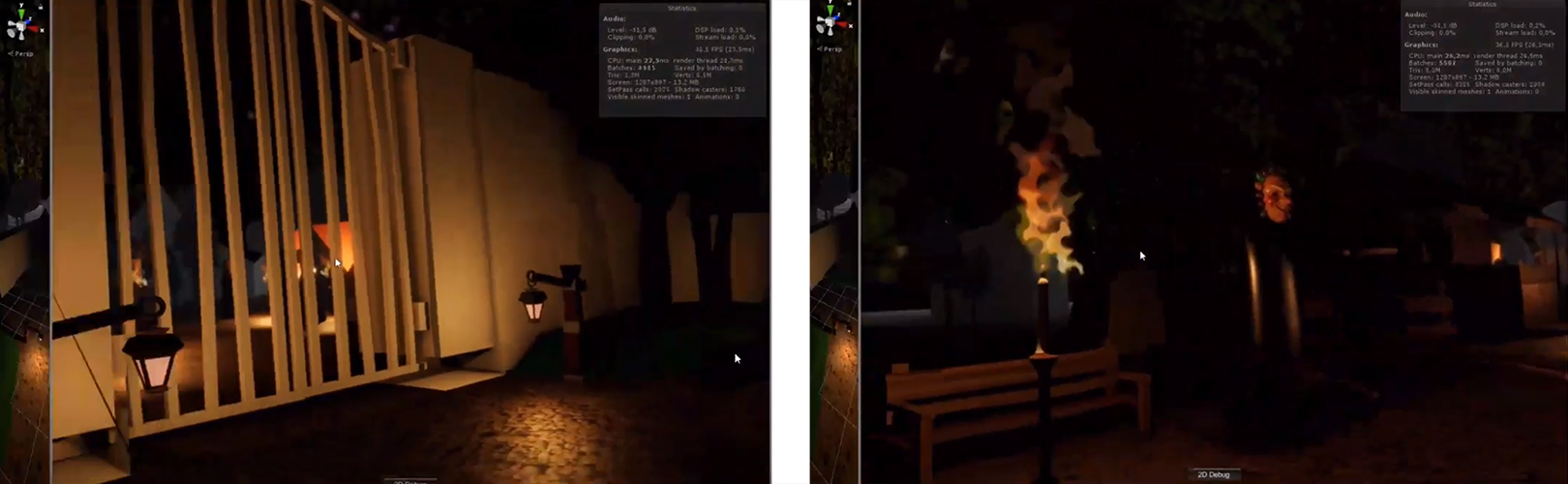}
\caption{Screenshots from the prototype made by team 1.} \label{criteria}
\end{figure*}

The winning project, created by team 1, was inspired by the atmosphere of traditional German outdoor festivals. This project was the most developed visually, relying on a visually-pleasing environment with atmospheric lightning, ambient music and custom models. The scene takes place in a night setting, in a park area where a festival is located. The player starts out in front of a gate leading into the festival grounds, surrounded by trees and warm lightning from torches and fireplaces. After interacting with the gate it opens, showing an alley with multiple tents and characters in fairytale-like costumes consisting of a cloak and a mask. Inside each tent there is a different minigame based on traditional childhood games widespread in Germany, such as Topfschlagen (heat-cold), Ein, Zwei, Drei... Halt! (Baba Yaga is watching) and Feuer-Wasser-Sturm-Blitz (fire-water-air-ground). The avatars instruct the player in German on how to proceed with each game; there are no subtitles to focus on improving listening competence. After finishing all games, all the NPCs move to a big bonfire located at the end of the alley, where an effigy is burned. The team received 49 points from the pre-determined evaluation criteria, as well as additional 3.5 points for graphics, 2.5 points for the idea, 1 for the mood, 1 for effects, and 0.5 for the potential.

\subsubsection{Second place}

\begin{figure*}
\includegraphics[width=\textwidth]{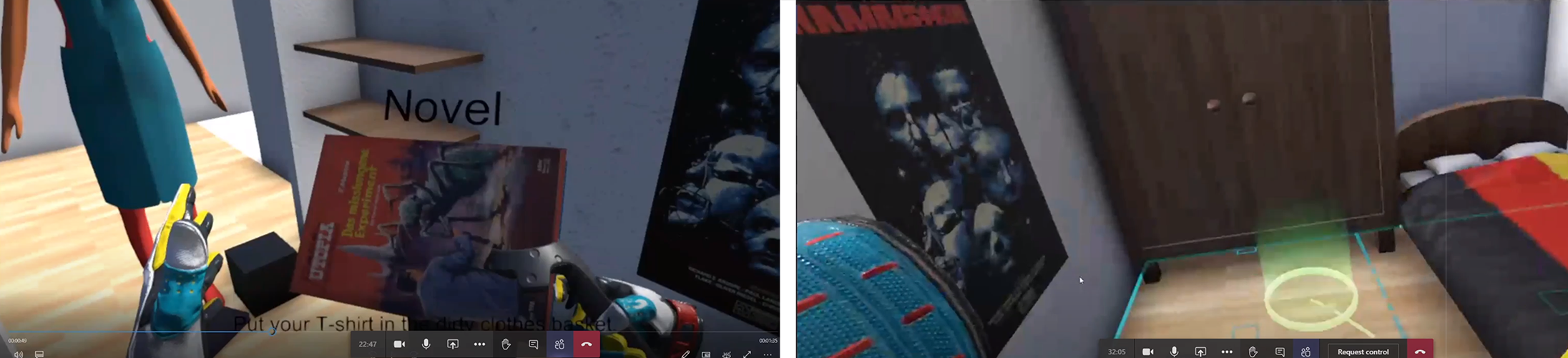}
\caption{Screenshots from the prototype made by team 3.} \label{criteria}
\end{figure*}

The premise used for this project (by the team number 3) was a situation when an angry German mother is forcing the user to perform household chores, shouting the instructions in German. The environment conveyed the local atmosphere in a humorous way, with the mother dressed in traditional German apparel and over-the-top decorations in the interior (such as bedsheets with the German national flag or Rammstein poster on the wall). When picking up each object, its name was displayed in German, allowing the user to check the vocabulary. The quality of performance influenced the behaviour of the mother avatar, who reacted verbally in response to the user’s interactions with objects, saying sentences such as “what are you doing?!” in the case of mistakes or “that’s my son!” when the task was done correctly. The environment resembled an apartment with multiple chores to perform in different rooms. In the bedroom, the user had to pick up objects from the floor and place it in appropriate locations, e.g. dirty clothing in a clothing basket. In the kitchen there were dirty dishes to wash using a sponge. The environment was fully functional and interactive, with features such as a script allowing dirty clothing to spawn in a randomized way. The team grounded the application in a mnemonic device of associating information with an unusual or weird situation, which facilitates memorizing. They also presented ideas for the future development of the app, listing possible additional functionalities and further improvements. The team received 37.5 points from the pre-determined evaluation criteria, as well as 1 for humour, 1 for functionalities, and 0.5 for potential.

\subsubsection{Third place}

\begin{figure*}
\includegraphics[width=\textwidth]{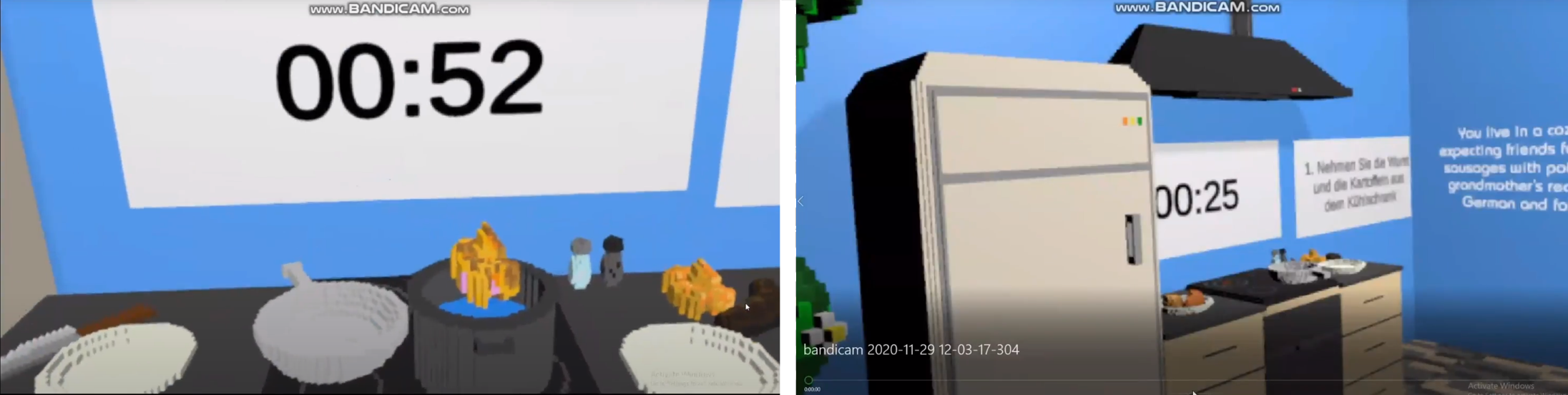}
\caption{Screenshots from the prototype made by team 4.} \label{criteria}
\end{figure*}

Team number 4 decided to make a game about cooking national German dishes (cooked sausage with potatoes). They created a scenario described as “You live in a cozy house near Munich, you’re expecting friends for dinner tonight. Cook the best sausages with potatoes according to your great-grandmother’s recipe”. They created a virtual kitchen in which the user had to accomplish the task of cooking traditional German sausages with potatoes according to instructions in German displayed above the kitchen worktop. Next to instructions there was a timer counting down the time left to finish each stage. Ingredients had to be found inside a fridge, put on a frying pan and into a pot (which needed to be filled with water from the tap) placed on the stove and served on a platter. Although it was not clearly visible in the prototype due to the time constraints, the environment was intended to imitate a traditionally furnished German kitchen and convey the atmosphere of local culture. The team intended to extend the simulation with other recipes from the German cuisine. The team received 29 points from the pre-determined evaluation criteria and no additional points.

\section{Discussion and Conclusions}

In this article we reported a case study of a Language and Culture-oriented VR hackathon with Goethe-Institut which was conducted online.  Overall, in comparison to our previous Hackathon, the projects presented during this edition were more advanced, allowing for greater interaction and immersion. We improved the formula of the event based on lessons learned from its previous edition and based on this experience we offer additional considerations:
\begin{enumerate}
\item First, we provided the participants with the \textbf{ability to add their own technology track} to mitigate one of the biggest barriers related to having specialized VR equipment at hand. Yet, in our case, granting the participants more freedom did not work. The team in this track, despite having suggested this track themselves, did not finish their project, as likely they lacked the starting point provided by a pre-defined framework and limiting technology options.
\item Next, for our default track we have created\textbf{ a custom VR framework, to serve as a starting point for the participants' designs to skip the repetitive early development stage}. Based on our pre-hackathon experience and questions we recommend to add a test-run of the said framework at least one day before the hackathon with a small "quest" to complete involving all the technical skills needed to take part in the hackathon. The results of the quest project would not count towards the hackathon scoring. We are also considering developing the framework further based on participant feedback, especially to include more ready-made scripts to enable a greater range of interactions, without limiting the participants' creativity by providing sets of assets directed at a certain interpretation of the tasks. In general, the use of programming frameworks is widespread in professional projects - therefore we recommend this as a good practice during hackathons.  Hackathon organizers, thanks to such provided framework, could better match the direction of participants' efforts to the hackathon goal. 
\item Finally, for this hackathon we outlined best modern research-backed language-learning practices and methods and gathered them into \textbf{actionable evaluation criteria to ensure better understanding of the goal}. Whenever expert knowledge and practices are expected to be exhibited by non-professionals, such as in this case philology and methodology studies, it is advisable to briefly and clearly state the scope and expectations derived from subject expertise. In our case, we evaluated our criteria against their performance and have come up with an improved set for this specific purpose. The best criteria were: engagement, culture, discovery, experiential, immersive and communicative as explained in Fig \ref{critnew} -- to these criteria we would add: graphics, innovative idea as well as potential for the project development.
\end{enumerate}

We recommend the steps of providing a starting framework and extending expert evaluation criteria, to encourage the creation of more advanced projects, as well as to improve the experience of hackathon participation for the teams - which in general thanks to these practices which allow them to focus their creativity within a smaller range of possibilities, produce much better results. \textbf{These are especially important for online events, where direct communication within teams as well as with experts and mentors may be more difficult - as such, then these solutions provide an easy and common reference point.}

\section*{Acknowledgments}

We would like to thank the many people and institutions gathered together by the distributed Living Lab Kobo and HASE Research Group (Human Aspects in Science and Engineering) for their support of this research. In particular, the authors would like to thank the members of XR Lab Polish-Japanese Academy of Information Technology and Emotion-Cognition Lab SWPS University as well as other HASE member institutions.

\bibliographystyle{splncs04}
\bibliography{bibliography}
\end{document}